\documentclass{article}
\usepackage[latin1]{inputenc}
\usepackage[english]{babel}
\usepackage{dsfont}
\usepackage[cyr]{aeguill}
\usepackage{amsmath,amssymb}
\usepackage{amsthm}
\usepackage{eurosym}
\usepackage{multicol}
\usepackage{color}
\usepackage{fancyhdr}
\usepackage[pdftex]{graphicx}
\usepackage[pdftex]{hyperref}
\usepackage{verbatim}
\usepackage{setspace}
\usepackage{stmaryrd}

\newcommand{\indic}{\mathds{1}} 
\newcommand{\E}{\mathbb{E}} 

\DeclareMathOperator*{\argmax}{argmax \ }
\DeclareMathOperator*{\argmin}{argmin \ }

\theoremstyle{plain}

\providecommand{\keywords}[1]{\textbf{\textit{Keywords --}} #1}

\title{Complexity measure, kernel density estimation, \\ bandwidth selection, and the efficient market hypothesis}

\author{Matthieu Garcin\footnote{ Léonard de Vinci Pôle Universitaire, Research center, 92916 Paris La Défense, France, matthieu.garcin@m4x.org.}} 

\date{\today}

\begin{document}

\maketitle

\begin{abstract}
We are interested in the nonparametric estimation of the probability density of price returns, using the kernel approach. The output of the method heavily relies on the selection of a bandwidth parameter. Many selection methods have been proposed in the statistical literature. We put forward an alternative selection method based on a criterion coming from information theory and from the physics of complex systems: the bandwidth to be selected maximizes a new measure of complexity, with the aim of avoiding both overfitting and underfitting. We review existing methods of bandwidth selection and show that they lead to contradictory conclusions regarding the complexity of the probability distribution of price returns. This has also some striking consequences in the evaluation of the relevance of the efficient market hypothesis. We apply these methods to real financial data, focusing on the Bitcoin.
\end{abstract}




\section{Introduction}

The efficient market hypothesis (EMH) stipulates that the current price of a financial asset reflects all the available information and so that one cannot make profitable forecasts in average. Reality of financial markets is often far from this theory, considering that statistical arbitrage is a lucrative business. More methodically, various statistical approaches may be used to determine whether a particular asset follows the EMH or not. Among them, one can cite the market information, inspired by the information theory and which relies on a particular statistic exploiting the discretized probability of price returns~\cite{Risso,MBM,Ducournau,BG}. The market information is precisely the difference between the Shannon entropies of two specific probability distributions. 

This notion of entropy is in addition often used as a measure of complexity. For this other reason, it is an instructive tool in finance. Indeed, financial markets are characterized by the interaction of many heterogeneous agents~\cite{CCM,Bouchaud,KK}: from the perspective of a physicist, it corresponds to the definition of a complex system. As a consequence, models in finance, such as those depicting the probability distribution of price returns, must reflect this complexity. 

The entropy-based market information relies on a discretized version of the probability density function (pdf) of price returns, in which one simply considers two states, namely positive and negative returns. This widespread simplification may be too simple to describe the complexity of the true pdf. Several attempts exist to finitely increase the number of states in this discretization~\cite{SLOS,LBA}. But, in this case, the corresponding entropy, which is still a complexity measure, cannot be directly used as a measure of market efficiency. This highlights the mismatch between complexity and efficiency: the two notions are of interest in finance, but each needs appropriate measuring tools.

A proper description of complexity may require infinitely increasing the number of states in the discretized probability distribution used in the entropy. This is then equivalent to working directly with the continuous pdf. Many statistical approaches are possible to estimate this pdf, keeping in mind the two main pitfalls of overfitting and underfitting. Overfitting arises when one uses a model very close to the empirical distribution, whereas underfitting often corresponds to the selection of a too simple parametric model. Since we don't have a preconceived idea about an appropriate parametric model, we prefer working with a nonparametric model, namely the kernel density estimation. This model relies on a key free parameter called bandwidth, which makes it possible to tune the smoothing of the distribution and thus to hopefully avoid both overfitting and underfitting. A very rich literature in statistics focuses on bandwidth selection~\cite{JMS,Tsybakov,WJ}. But existing methods are not based on an analysis of complexity, whereas both measures of complexity and of market efficiency strongly depend on this bandwidth parameter. From a model risk perspective, it is hard to accept that the conclusions regarding for instance the efficiency of a particular asset vary according to the methodological choice of a bandwidth and not only according to the data.

The above remarks show that it is important to study together complexity and kernel density estimation. Three main questions indeed arise. How can one define a complexity measure adapted to kernel densities? To which extent do complexity and market efficiency measures depend on the bandwidth parameter? How can one define a new bandwidth selection method based on the notion of complexity?

To our knowledge, the joint study of complexity and kernel approaches has not been the subject of much investigation. In the framework of nonlinear regression with unknown distribution of the residuals, we can cite an approach in which the parameter of the regression is estimated by the minimization of the entropy of the distribution, which is itself estimated by a kernel approach~\cite{WTP}. But it is the complexity in relation to the parameter of the regression, and not in relation to the bandwidth, which is put forward there. In another interesting contribution, a bandwidth selection method is proposed in order to minimize the uncertainty of the resulting kernel pdf. More precisely, the authors select the bandwidth minimizing the entropy of the estimated density~\cite{JHD}, approximated for the ease of the calculations by the re-substitution entropy~\cite{HLW}. In other words, this method provides the kernel density which deviates the most from the uniform distribution. 

The idea of interpreting complexity as the divergence from a reference distribution also appears in the Lopez-Ruiz, Mancini, and Calbet (LMC) complexity. It is a widespread statistic used to measure the complexity of any probability distribution, even not based on the kernel approach~\cite{LMC}. It combines Shannon's entropy with a measure of divergence of the pdf with respect to the uniform distribution. In the LMC approach, the two states with the lowest complexity correspond to the uniform and the Dirac distributions. One can thus define the complexity as a divergence from a particular reference set of simplistic distributions. The selection of this reference set is to be made carefully and in line with the specificities of kernel densities and of the application field, in our case finance. It is one of the goals of this work.

In Section 2, we expose the kernel density estimation as well as the LMC complexity, and we propose an alternative complexity measure adapted to our financial framework. Based on this measure, we introduce a new bandwidth selection method for kernel density estimation. A simulation study, in Section 3, shows how this method behaves with respect to other existing ones. In Section 4, we propose an application to financial data, more precisely Bitcoin, as well as an empirical study of the link between bandwidth selection and market efficiency. Section 5 concludes.

\section{Complexity of kernel densities}

In this section, we present how nonparametric density estimation articulates with complexity. We start by an introduction to kernel density estimation and bandwidth selection. Then we focus on complexity measures and particularly on one consistent with kernel densities. We then deduce a complexity-based method of bandwidth selection. 

\subsection{Kernel density estimation}\label{sec:KDE}

We assume we are given $n$ independent and identically distributed (i.i.d.) observations $X_1,...,X_n$, of pdf $f$. The kernel density is defined, for $x\in\mathbb R$, by
\begin{equation}\label{eq:kde_pdf}
\widehat f_h(x)=\frac{1}{nh}\sum_{i=1}^{n}{K\left(\frac{x-X_i}{h}\right)},
\end{equation}
where $h>0$ is the bandwidth and $K$ is the kernel, a function following the same rules as a pdf, namely it is positive, integrable and its integral is one~\cite{WJ,Tsybakov}. The corresponding cumulative distribution function (cdf) is
\begin{equation}\label{eq:kde_cdf}
\widehat F_h(x)=\frac{1}{n}\sum_{i=1}^{n}{\mathcal K\left(\frac{x-X_i}{h}\right)},
\end{equation}
where $\mathcal K(x)=\int_{-\infty}^xK(y)dy$. In Sections~\ref{sec:simul} and~\ref{sec:applifinance}, we focus on the particular case of a Gaussian kernel. The quality of the estimation does not much depend on the choice of a specific function for $K$. The impact of the selected bandwidth has much greater consequences: the larger $h$, the smoother the estimated density.

The literature about bandwidth selection is very large. Whatever the method, the common goal is often the minimization of an accuracy criterion, the mean integrated squared error (MISE): 
$$\text{MISE}=\E\left[\int_{\mathbb R}\left(\widehat f_h(x)-f(x)\right)^2dx\right].$$
When $n\rightarrow\infty$, an asymptotic expansion simplifies the expression of MISE, which leads to the asymptotic mean integrated squared error (AMISE), a linear function of $h^{-1}$ and $h^4$. As a consequence, we have a straightforward minimizer of AMISE~\cite{Silverman,JMS}:
\begin{equation}\label{eq:hAMISE}
h_{\text{AMISE}}=\left[\frac{\int_{\mathbb R}K(x)^2dx}{n(\int_{\mathbb R}f''(x)^2dx)(\int_{\mathbb R}x^2K(x)dx)^2}\right]^{1/5}.
\end{equation}
Unfortunately, this expression of the asymptotically optimal bandwidth relies on $f$, which is unknown. We can bypass this difficulty by using for instance a cross-validation or a plug-in estimator. The plug-in consists in replacing the true and unknown density $f$ in equation~\eqref{eq:hAMISE} by its estimator $\widehat f_h$. It transforms equation~\eqref{eq:hAMISE} as a new fixed-point equation in $h$, which can easily be solved numerically.

Beyond this classical approach, we can cite other methods based on other criteria than MISE. For instance, the literature on time-varying densities proposes some criteria based instead on the maximization of the likelihood of the observations~\cite{HO}, or on a validation technique of the pdf using the probability integral transform (PIT) approach, which states under some conditions that PITs, that is $\widehat F_h(X_i)$, must be i.i.d. uniform variables~\cite{GKL,AG}. These methods have been extended to static densities and are thus alternatives to the traditional MISE-based methods~\cite{GKL}. They rely on a simple train-validation approach: in addition to the $n$ observations $X_1,...,X_n$, we are also given an i.i.d. validation set $X_{n+1},...,X_{n+m}$. The likelihood-based bandwidth is the solution of
$$h_{\text{lik}}=\underset{h>0}{\argmax} \sum_{i=1}^{m}{\log\left(\widehat f_{h}(X_{n+i})\right)}$$
and the PIT-based bandwidth the solution of
$$h_{\text{PIT}} = \underset{h>0}{\argmin}\ \underset{0\leq\tau\leq\nu}{\max}\left(\sqrt{m-\tau}\times k_{\tau}\left(\widehat F_h(X_{n+1}),...,\widehat F_h(X_{n+m})\right)\right),$$
where in both methods the pdf $\widehat f_{h}$ and the cdf $\widehat F_{h}$ are estimated on $X_1,...,X_n$ only, following equations~\eqref{eq:kde_pdf} and~\eqref{eq:kde_cdf}, where $\nu$ is a free parameter defining the time range in which the independence of the PITs is taken into account in the criterion,\footnote{ In Section~\ref{sec:simul}, we consider $\nu=22$ and $m=n$.} and where $k_{\tau}$ is the Kolmogorov-Smirnov divergence statistic between the theoretical distribution of an independent pair of uniform variables and the empirical distribution of all the pairs of PITs separated by a lag $\tau$:
$$k_{\tau}(z_1,...,z_m)=\left\{\begin{array}{ll}
\underset{1\leq i\leq m}{\max}\left|z_i-\frac{1}{m+1}\sum_{j=1}^{m}{\indic_{z_j\in[0,z_i]}}\right| & \text{if } \tau=0 \\
\underset{1\leq i\leq m-\tau}{\max}\left|z_iz_{i+\tau}-\frac{1}{m-\tau+1}\sum_{j=1}^{m-\tau}{\indic_{z_j\in[0,z_i]}\indic_{z_{j+\tau}\in[0,z_{i+\tau}]}}\right| & \text{else}.
\end{array}\right.$$

\subsection{Complexity measures}

The complexity measure put forward by the information theory is Shannon's entropy. According to this metric, the least complex probability distribution is the degenerate one in which all the probability rests on a single state, also known as Dirac distribution. At the opposite, the uniform distribution is the most complex, in the sense that it reflects the highest possible uncertainty. 

Nevertheless, the uniform distribution belongs to the family of very standard distributions, so that its complexity in an informational perspective does not correspond to a complexity in terms of modeling. For this reason, a common practice in physics consists in using instead the LMC measure~\cite{LMC}. Indeed, in physics, one can think about two basic examples of simple systems, for which the complexity should thus be equal to zero: the perfect crystal and the isolated ideal gas. The perfect crystal has a zero entropy, because a small piece of information is enough to describe the whole system. The isolated ideal gas is totally disordered and so each state has the same probability. These two systems correspond respectively to the degenerate distribution described above and to the uniform distribution. The sole entropy is then not an appropriate complexity measure: the low complexity is to be defined not only by the maximum order but also by the maximum disorder. Therefore, the LMC complexity applied to a pdf $g:[a,b]\rightarrow\mathbb R$ is the product of its Shannon's entropy $H(g)$ with $D(g)$, an Euclidean divergence statistic of $g$ from the uniform distribution:
$$\mathcal C_{\text{LMC}}(g)=H(g)\times D(g),$$
where
$$H(g)=-\int_{a}^b g(x)\log(g(x))dx \qquad \text{and} \qquad D(g)=\int_{a}^b\left(g(x)-\frac{1}{b-a}\right)^2dx.$$
It is worth noting that generalizations of the LMC complexity exist in which $H(g)$ is another entropy, like Tsallis, escort-Tsallis, or Renyi entropies, and $D(g)$ is another divergence statistic, like Wootters distance, relative Kullback entropy, or Jensen divergence~\cite{RML}.

Going back to the kernel density estimation, $\mathcal C_{\text{LMC}}(\widehat f_h)$ depends on $h$ in a trivial way. Indeed, unless the true pdf $f$ is close to the degenerate ``perfect-crystal'' pdf, $\widehat f_h$ is itself very far from this degenerate distribution. It can also only approach the uniform pdf when one increases $h$. The highest complexity would thus correspond to $h\rightarrow 0$ and the lowest to $h\rightarrow\infty$. The selection of the bandwidth as a maximizer of the LMC complexity criterion is thus trivial and far from the more realistic solutions proposed in Section~\ref{sec:KDE}, from the statistician's perspective.

\subsection{Appropriate complexity measure of a kernel density}\label{sec:AppropriateComplexity}

Following the LMC idea of defining two simple states of low complexity, namely maximum order and maximum disorder, we propose a new complexity measure which would be relevant for kernel densities in a context of financial data. 

In this framework, we consider that the maximum disorder is the empirical distribution, towards which the kernel distribution tends when $h\rightarrow 0$. In this empirical setting, the distribution is overconfidently supposed to be perfectly described by the limited number of observations at hand and the estimated density is thus the best example of overfitting. 

At the opposite, the maximum order corresponds to a simplistic parametric distribution. In the LMC complexity, this distribution is the uniform one, but such a distribution is not relevant in finance, as soon as no one would use it to approximate the pdf of price returns. Instead, the simple parametric distribution which is overused in finance is the Gaussian distribution. We will thus naturally consider that distributions close to the Gaussian one have a low complexity. Since reality is often more complex than the Gaussian distribution and deserves a more sophisticated model, this distribution is an example of underfitting.

Consequently, we define our complexity measure $\mathcal C_h$, of a kernel density estimate of bandwidth $h$, by
\begin{equation}\label{eq:NewComplexity}
\mathcal C_h=\rho(\mathcal E_h,\mathcal P_h),
\end{equation}
where $\mathcal E_h$ is the divergence of $\widehat F_h$ from the empirical distribution, $\mathcal P_h$ is the divergence of $\widehat f_h$ from the parametric (in our case Gaussian) distribution, and $(x,y)\in[0,\infty)^2\mapsto\rho(x,y)$ is an aggregation function. 

Many divergence statistics may be used for $\mathcal E_h$ and $\mathcal P_h$, like the Kolmogorov-Smirnov statistic, Hellinger distance, Wasserstein distance, or Kullback-Leibler divergence~\cite{TGG,GKL}. One of the difficulties is that, in $\mathcal E_h$, we want to determine the divergence between a continuous and a discrete distribution. We can thus either transform the continuous pdf in a discrete probability thanks to binning, or we can work directly with the cdf of each distribution instead. More specifically, in the application of the method in Sections~\ref{sec:simul} and~\ref{sec:applifinance}, we focus on the Kolmogorov-Smirnov statistic:
$$\mathcal E_h=\underset{x\in\mathbb R}{\sup}\left|\widehat F_{h}(x)-\frac{1}{n}\sum_{i=1}^n \indic_{X_i\leq x}\right|=\underset{j\in\llbracket 1,n\rrbracket}{\max}\left(\max\left(\left|\widehat F_{h}(X_j)-\frac{1}{n}\sum_{i=1}^n \indic_{X_i\leq X_j}\right|,\left|\widehat F_{h}(X_j)-\frac{1}{n}\sum_{i=1}^n \indic_{X_i<X_j}\right|\right)\right).$$
Regarding $\mathcal P_h$, we select the Kullback-Leibler divergence because we want to keep the information-theoretic flavour of the LMC complexity. However, the traditional Kullback-Leibler divergence requires a very fine discretization of the integral to lead to accurate results. We thus prefer an extension of this statistic in which each pdf is replaced by the corresponding cdf, namely the cumulative Kullback-Leibler divergence~\cite{PRS,DCL}:
$$\mathcal P_h=\int_{\mathbb R}{\widehat F_{h}(x)\log\left(\frac{\widehat F_{h}(x)}{G_{\widehat \theta}(x)}\right)dx},$$
where $G_{\theta}$ is the Gaussian cdf of parameter vector $\theta$ and $\widehat{\theta}$ is the parameter estimated on the dataset $X_1,...,X_n$ using the maximum-likelihood approach.

Though the complexity of the kernel density defined by equation~\eqref{eq:NewComplexity} tends toward zero when $h\rightarrow 0$, because of the proximity of $\widehat F_{h}$ with the empirical distribution, the quantity $\mathcal P_h$ will never be equal to zero for this kind of nonparametric distribution. At best, we can find the bandwidth $h_p$ which leads to the kernel density closest to $G_{\widehat \theta}$:
$$h_p=\underset{h>0}{\argmin} \mathcal P_h.$$
We focus on values of bandwidth in the interval $(0,h_p]$, because increasing $h$ over $h_p$ will lead to a density smoother than $\widehat f_{h_p}$, which we consider to be the closest to the least sophisticated parametric solution, that is the closest to the smoothest conceivable pdf.

Regarding the aggregation function $\rho$, it must have a positive value, be equal to zero if and only if the tested distribution is the empirical or the Gaussian one, be symmetrical in its two arguments, and be a nondecreasing function of each of its arguments. Instead of the product function used in LMC complexity, we use the minimum function between the scaled divergences. The purpose of the scaling is to make a fair comparison between the two divergences $\mathcal E_h$ and $\mathcal P_h$, by dividing them by their maximum in the acceptable interval of bandwidths $(0,h_p]$:
\begin{equation}\label{eq:NewComplexityScaled}
\mathcal C_h=\min\left(\frac{\mathcal E_h}{\underset{\eta\in(0,h_p]}{\max}\mathcal E_{\eta}},\frac{\mathcal P_h}{\underset{\eta\in(0,h_p]}{\max}\mathcal P_{\eta}}\right).
\end{equation}

\subsection{A complexity-based method of bandwidth selection}\label{sec:NewBandwidth}

As exposed in Section~\ref{sec:AppropriateComplexity}, we focus on values of bandwidth in the interval $(0,h_p]$. We consider that bandwidths higher than $h_p$ lead to overly smoothed densities. A good illustration of the usefulness of this limitation is provided by the case $h\rightarrow\infty$. In this situation, the kernel density will be of the same type as the kernel function itself. In other words, with a Gaussian kernel, $\widehat f_{h}$ will then tend toward a Gaussian pdf. But this Gaussian pdf is absurd: it is very far from $G_{\widehat \theta}$ because of its largely overestimated variance parameter.

We propose a bandwidth selection method based on the complexity put forward in equation~\eqref{eq:NewComplexity}. This complexity has been built so as to reach low values in case of overfitting and of underfitting, namely for $h$ close to 0 and for $h$ close to $h_p$. Therefore, an appropriate fitting of the true density should be a compromise between these two extreme cases and should therefore correspond to a higher value of $\mathcal C_h$, more consistent with the complex nature of reality. We thus propose to select the bandwidth $h_c$ which maximizes the complexity in the acceptable interval of bandwidths $(0,h_p]$:
\begin{equation}\label{eq:MaximumComplexityBandwidth}
h_c=\underset{h\in(0,h_p]}{\argmax} \mathcal C_h.
\end{equation}
This complexity criterion to be maximized, insofar as it is intended to avoid underfitting and overfitting, follows the same goal as information criteria to be minimized, like AIC or BIC, which one uses to select an appropriate parametric model.
 
The next sections study on simulations and on real data the outcome of this new bandwidth selection method.

\section{A simulation study}\label{sec:simul}

We now study the dependence of the complexity, as defined in equation~\eqref{eq:NewComplexityScaled}, on the value of the bandwidth. In particular, we compare the complexity obtained for $h_{\text{AMISE}}$, $h_{\text{PIT}}$, $h_{\text{lik}}$, and for the complexity-based bandwidth $h_c$ introduced in Section~\ref{sec:NewBandwidth}. We work with simulated datasets of size 1,000. For $h_{\text{PIT}}$ and $h_{\text{lik}}$, as exposed in Section~\ref{sec:KDE}, we also consider 1,000 additional observations in the same probability law for the validation. The simulations are generated according either to a standard Gaussian distribution, or to a mixture of two Gaussian distributions $0.6g(x+1.25)+0.4g(x-1.25)$, where $g$ is the standard Gaussian density, or to a fat-tailed Student distribution with 5 degrees of freedom. 

Table~\ref{tab:bandwidthSimul} displays for each of these three simulated datasets the optimal bandwidth, following the four methods exposed above. For the Gaussian and the Student distributions, $h_{\text{PIT}}$ is the smallest bandwidth. Then, the values of $h_c$ and $h_{\text{AMISE}}$ are quite close to each other, whereas $h_{\text{lik}}$ is the largest bandwidth. The two methods extended from the dynamic framework, $h_{\text{PIT}}$ and $h_{\text{lik}}$, thus seem to have a singular behaviour when applied to our static case, with respect to more classical approaches like AMISE or to the complexity-based approach. Table~\ref{tab:bandwidthSimul} also reports the complexity corresponding to the kernel density of each given bandwidth. As expected by the definition of $h_c$, the complexity-based bandwidth leads to a density with a higher complexity than the one estimated with $h_{\text{AMISE}}$ and even more so with $h_{\text{PIT}}$. But the complexity related to $h_{\text{lik}}$ is even higher than $\mathcal C_{h_c}$. However, this apparent paradox does not contradict the definition of $h_c$, because $h_{\text{lik}}$ is greater than $h_p$, that is it is out of the acceptable interval of bandwidths. In other words, the selection of $h_{\text{lik}}$ leads to an excessive smoothing of the kernel density.

\begin{table}[htbp]
\centering
\begin{tabular}{|l|c|c|c|c|c|}
\hline
Bandwidth & Gaussian & Gaussian mixture & Student \\
\hline
 $h_c\ (\mathcal C_{h_c})$ &  0.106 (0.71) &  0.413 (0.68) &  0.266 (0.91) \\
 $h_{\text{AMISE}}\ (\mathcal C_{h_{\text{AMISE}}})$ &  0.119 (0.67) &  0.320 (0.51) &  0.201 (0.74) \\
 $h_{\text{PIT}}\ (\mathcal C_{h_{\text{PIT}}})$ &  0.090 (0.65) &  0.164 (0.39) &  0.055 (0.42) \\
 $h_{\text{lik}}\ (\mathcal C_{h_{\text{lik}}})$ &  0.323 (1.54) &  0.308 (0.50) &  0.473 (1.00) \\
 $h_p\ (\mathcal C_{h_p})$ &  0.182 (0.58) &  0.550 (0.59) &  0.315 (0.90) \\
\hline
\end{tabular}
\begin{minipage}{0.9\textwidth}\caption{Bandwidth and corresponding complexity of the kernel density for the four selection methods and the border value $h_p$, for each of the simulated datasets.}
\label{tab:bandwidthSimul}
\end{minipage}
\end{table}

When working with a two-mode distribution, namely the Gaussian mixture, we get some different results. Table~\ref{tab:bandwidthSimul} shows that $h_c$ is the greatest bandwidth and that $h_{\text{lik}}$ is much smaller than $h_p$ and thus does not overly smooth the density. 

\begin{figure}[htb]
	\centering
		\includegraphics[width=0.45\textwidth]{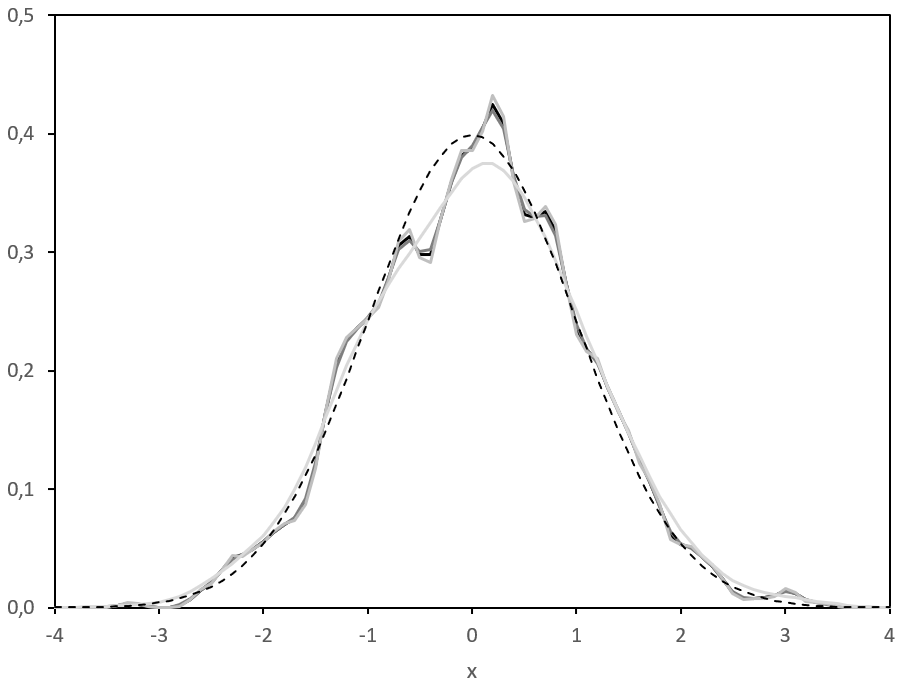} 
		\includegraphics[width=0.45\textwidth]{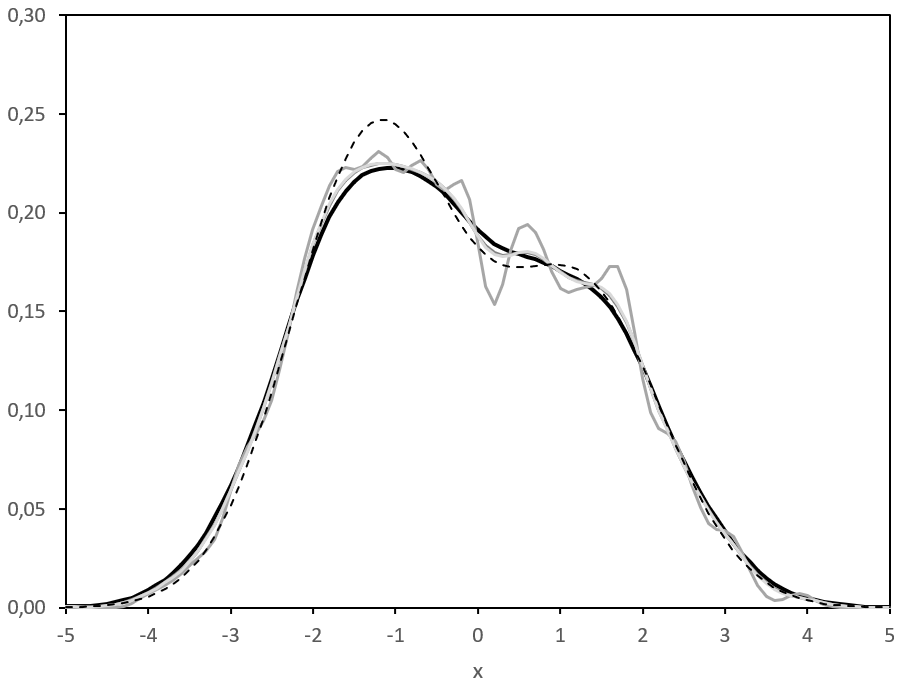} \\
		\includegraphics[width=0.45\textwidth]{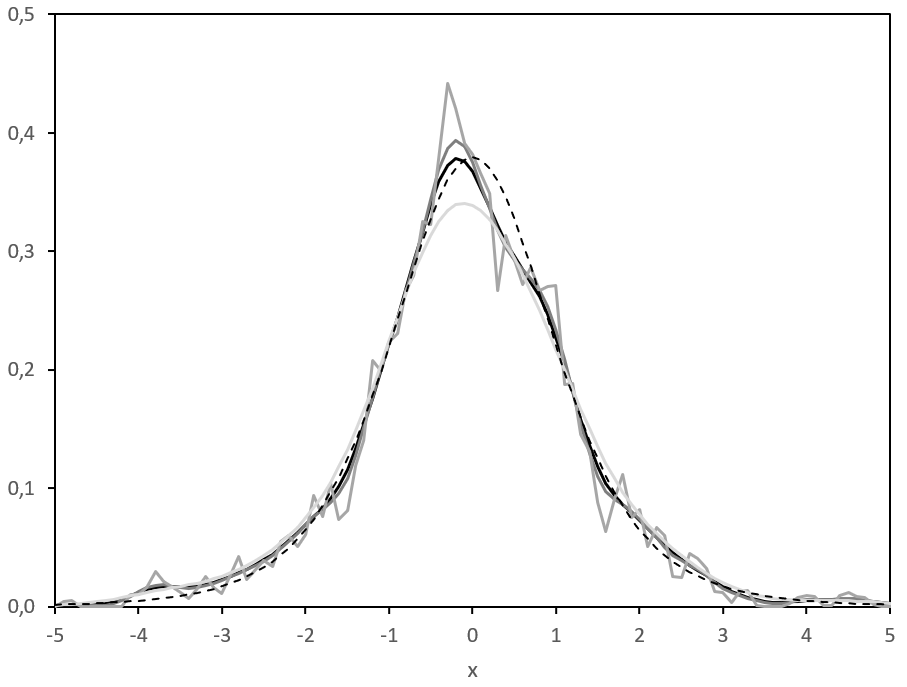} 
		\includegraphics[width=0.45\textwidth]{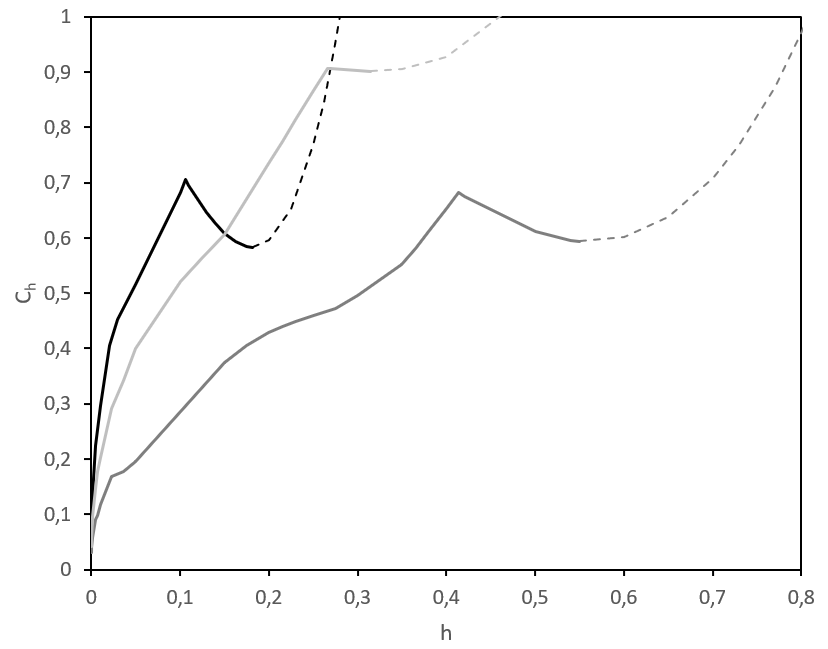} 
\begin{minipage}{0.9\textwidth}\caption{Top left and right and bottom left: true (dotted line), and estimated densities, with $h_c$, $h_{\text{AMISE}}$, $h_{\text{PIT}}$, $h_{\text{lik}}$ (from the darkest to the lightest), for the Gaussian, Gaussian mixture, and Student simulations. Bottom right: complexity with respect to the bandwidth for the Gaussian (black), Gaussian mixture (dark grey), and Student (light grey) simulations, the curves starting to be dotted at the bandwidth $h_p$.}
	\label{fig:simul}
\end{minipage}
\end{figure}

Figure~\ref{fig:simul} shows the estimated and the true densities, for the three datasets. One can for instance easily recognize the density estimated with  $h_{\text{PIT}}$, which always leads to a quite erratic curve. In some cases, the limited difference between two bandwidth leads to almost undistinguishable estimated densities. The complexity graph in Figure~\ref{fig:simul} also shows the dependence of $\mathcal C_h$ on $h$. Starting to a value asymptotically equal to 0 when $h\rightarrow 0$, increasing $h$ increases the complexity up to a very clear local maximum in $h_c$, before a decrease until the bandwidth reaches the value of $h_p$. Beyond this value, the complexity increases again, even above the value of $\mathcal C_{h_c}$, but this does not correspond to acceptable values of bandwidth as exposed in Section~\ref{sec:AppropriateComplexity}.

\section{Application to financial data}\label{sec:applifinance}

We now analyse the distribution of daily returns of a cryptocurrency. In particular, we describe how the bandwidth in the kernel density estimation impacts the complexity as well as the estimated market efficiency.

\subsection{Complexity of a cryptocurrency}

We study the Bitcoin (BTC-USD) in the time interval 2015-2022. We have imported the data from Yahoo Finance. In our analysis, we work with daily price returns and are interested in their kernel density estimation, limited to a one-year period, so that we get one estimated density for each of the 8 years in the time interval of the study. Since the Bitcoin market does never close, we have 365 observations each year, or 366 for leap years.

\begin{figure}[htb]
	\centering
		\includegraphics[width=0.45\textwidth]{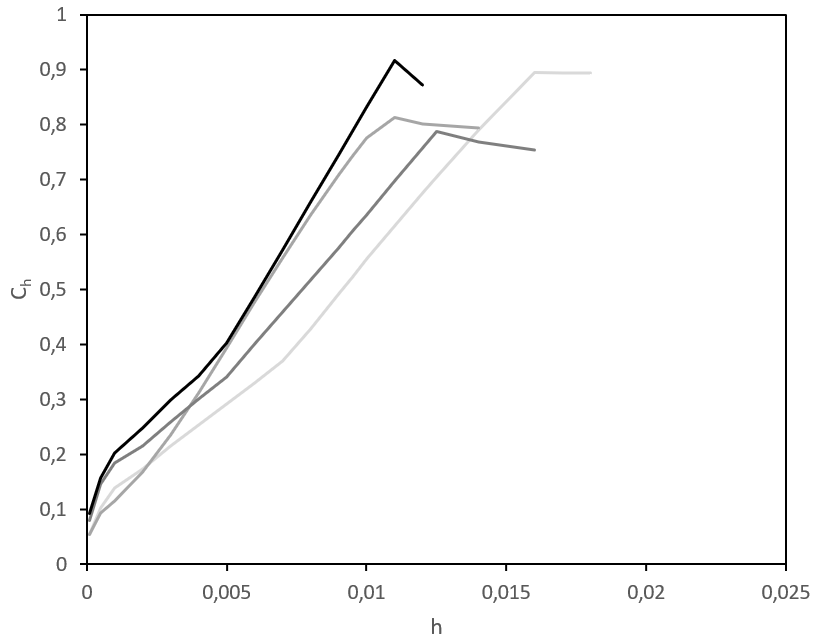} 
		\includegraphics[width=0.45\textwidth]{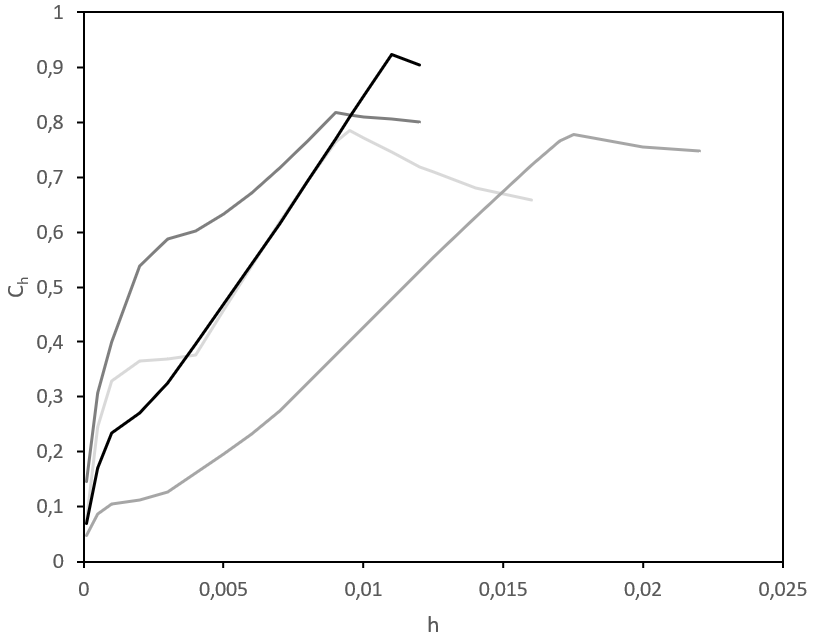}
\begin{minipage}{0.9\textwidth}\caption{Complexity $\mathcal C_h$ of the annual pdf of daily price returns for BTC-USD, estimated with a kernel approach, with respect to the bandwidth $h$ between 0 and the specific $h_p$ calculated for each density. The left graph is between 2015 and 2018 (from the lightest to the darkest curve, one curve per year) and the right graph between 2019 and 2022 (from the lightest to the darkest).}
	\label{fig:complexityBTC}
\end{minipage}
\end{figure}

\begin{figure}[htb]
	\centering
		\includegraphics[width=0.45\textwidth]{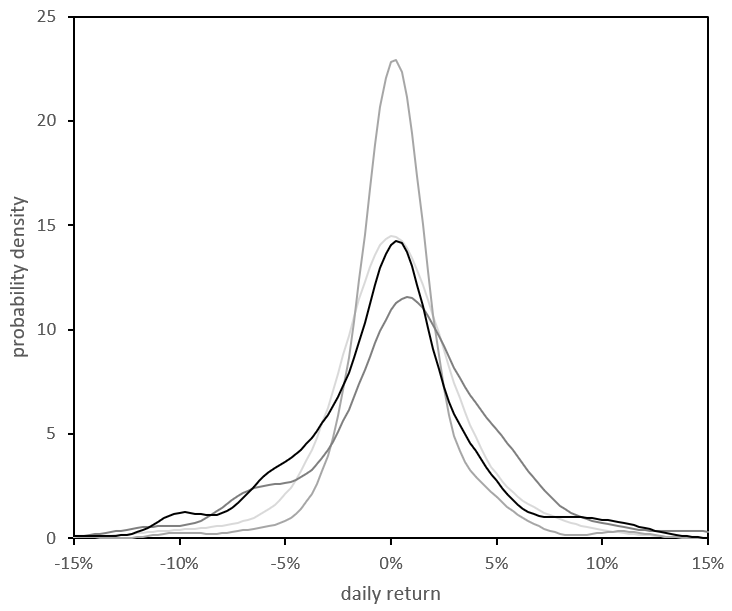} 
		\includegraphics[width=0.45\textwidth]{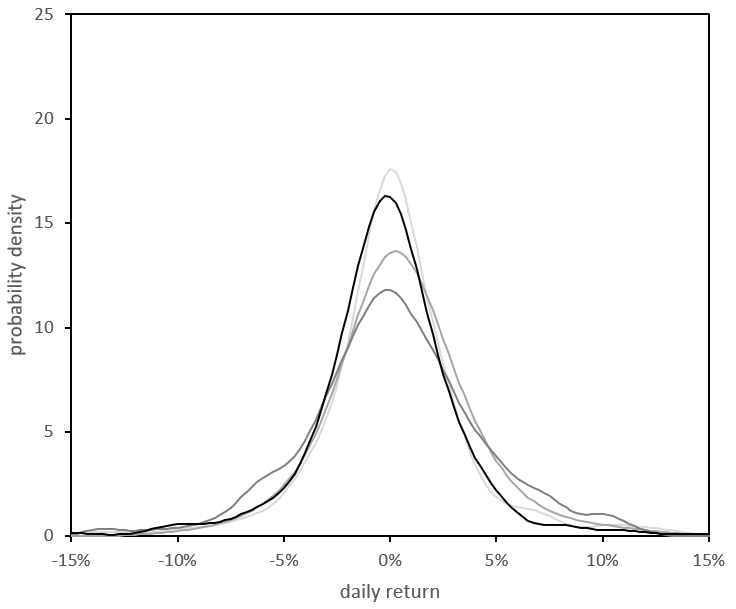}
\begin{minipage}{0.9\textwidth}\caption{Annual pdf of daily price returns for BTC-USD, estimated with a kernel approach and the bandwidth $h_c$ specified in Table~\ref{tab:bandwidthBTC}. The left graph is between 2015 and 2018 (from the lightest to the darkest curve, one curve per year) and the right graph between 2019 and 2022 (from the lightest to the darkest).}
	\label{fig:densityBTC}
\end{minipage}
\end{figure}

Table~\ref{tab:bandwidthBTC} displays the maximum-complexity bandwidth $h_c$ which we introduced in equation~\eqref{eq:MaximumComplexityBandwidth}. Depending on the year, this bandwidth is close to 0.010, but with two years for which the method requires a strongly higher smoothing: 2015 and 2020. Figure~\ref{fig:complexityBTC} shows the complexity $\mathcal C_h$ with respect to the bandwidth, for each year. By definition, the maximum of the curve is reached for $h=h_c$. Finally, the kernel density with the bandwidth $h_c$ for each year is in Figure~\ref{fig:densityBTC}. The shape of the density strongly depends on the study year, with noticeable differences even for years with two close bandwidths. For instance, in 2018, the density is skewed and has bumps far from its main mode. On the other hand, in 2019 the density is almost symmetrical with thin tails.

\begin{table}[htbp]
\centering
\begin{tabular}{|l|c|}
\hline
Year & $h_c$ \\
\hline
2015 & 0.016 \\
2016 & 0.011 \\
2017 & 0.013 \\
2018 & 0.011 \\
2019 & 0.010 \\
2020 & 0.018 \\
2021 & 0.009 \\
2022 & 0.011 \\
\hline
\end{tabular}
\begin{minipage}{0.9\textwidth}\caption{Bandwidth maximizing the complexity of the kernel density of daily price returns for the BTC-USD, by year.}
\label{tab:bandwidthBTC}
\end{minipage}
\end{table}

\subsection{Complexity and market efficiency}

The previous paragraph showed how the complexity may play a role in the bandwidth selection and so in the kernel density estimation. Of course, since the resulting density strongly depends on the bandwidth, every statistic based on this estimated density will be impacted by the selection method of the bandwidth. In particular, we now show how different bandwidths can lead to opposite conclusions regarding the market efficiency. We also note that the complexity of the distribution itself is not an appropriate measure of market efficiency and we need other statistics, using or not the estimated density, such as the probability of positive price returns, the entropy-based market information, or the Hurst exponent, which we all expose below.

\begin{figure}[htb]
	\centering
		\includegraphics[width=0.4\textwidth]{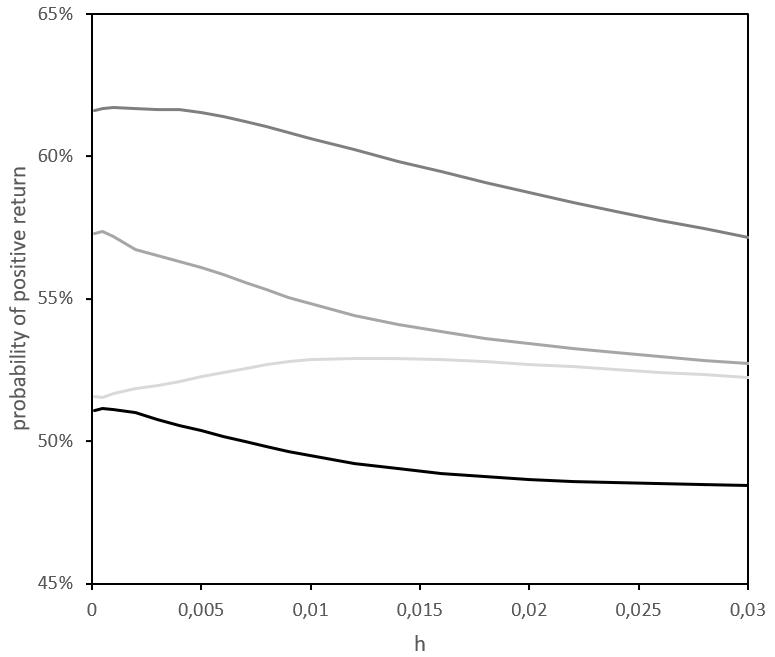} 
		\includegraphics[width=0.4\textwidth]{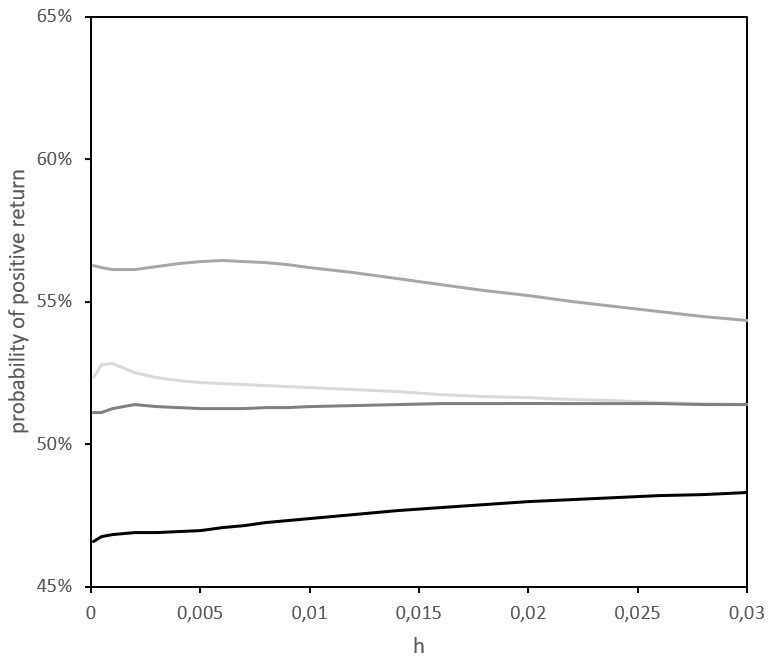}
\begin{minipage}{0.9\textwidth}\caption{Probability of a positive daily price return according to the estimated kernel density for BTC-USD, with respect to the bandwidth $h$. The left graph is between 2015 and 2018 (from the lightest to the darkest curve, one curve per year) and the right graph between 2019 and 2022 (from the lightest to the darkest).}
	\label{fig:ProbaHausseBTC}
\end{minipage}
\end{figure}

A first valuable statistic for measuring the market efficiency is simply the probability to have a positive price return, which should be $50\%$ according to the EMH. This probability can be obtained from the integration of the kernel density and it thus depends on the bandwidth. Figure~\ref{fig:ProbaHausseBTC} illustrates this idea. On the left of each curve, that is for $h\rightarrow 0$, we observe the empirical probability of positive price return. This value of course depends on the year and ranges between $46\%$ and $62\%$. Increasing the value of $h$ modifies this estimated probability. It can be either toward the value of $50\%$, that is toward more market efficiency (2016, 2017, 2020, 2022), or the opposite (2015), or even it can go from a value above $50\%$ to a value below $50\%$ (2018).

\begin{figure}[htb]
	\centering
		\includegraphics[width=0.45\textwidth]{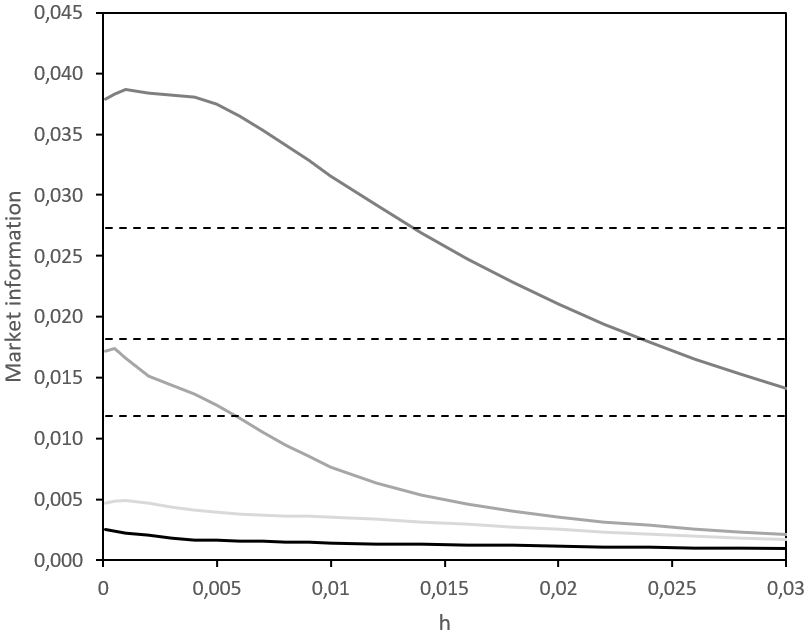} 
		\includegraphics[width=0.45\textwidth]{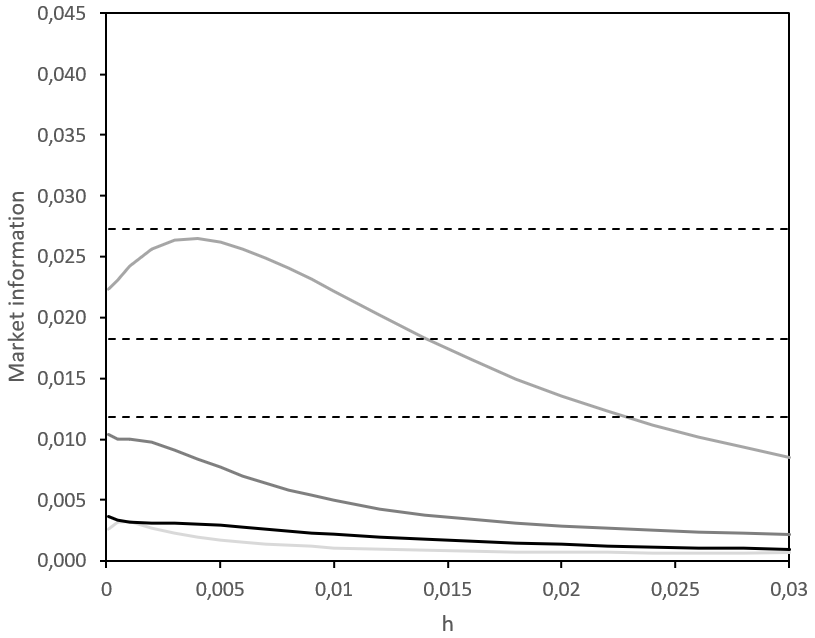}
\begin{minipage}{0.9\textwidth}\caption{Market information with one time lag, $I_h$, as defined in equation~\eqref{eq:MarketInfo}, calculated with the estimated kernel density of daily price returns for BTC-USD, with respect to the bandwidth $h$. The left graph is between 2015 and 2018 (from the lightest to the darkest curve, one curve per year) and the right graph between 2019 and 2022 (from the lightest to the darkest). The three dotted lines correspond to the confidence intervals of a zero market information with probability $95\%$, $99\%$, and $99.9\%$.}
	\label{fig:MarketInfoBTC}
\end{minipage}
\end{figure}

This first analysis makes it hard to conclude firmly about the presence or not of market efficiency, because of the simplicity of the statistic used and of its dependence on the bandwidth. We thus propose a second statistic to measure the market efficiency, namely methods based on information theory and in particular on Shannon's entropy. Mainly two approaches exist in this perspective, depending on the probability distribution used. If the distribution is the one corresponding to a specific permutation of the time series of price returns, we get the permutation entropy~\cite{BPompe}. We prefer working more directly with the distribution of price returns, which we have estimated with the kernel method. In this case, we get the market information~\cite{Risso,BG}. Following this approach, we simplify the dataset using a symbolic representation: a positive price return is noted 1, a negative price return 0. We want to determine whether the knowledge of a sequence of $L$ consecutive symbols is useful in forecasting the next one. In this work, we focus on the simplest case in which $L=1$. Using a kernel approach, with a given bandwidth $h$, we estimate the probability $p_{1,h}$ (respectively $p_{0,h}$) to have a positive (resp. negative) daily price return and the probability $\pi_{1,h}$ (respectively $\pi_{0,h}$) to have a positive price return conditionally on the fact that the previous price return is positive (resp. negative). We can then calculate the difference between the entropy of the distribution of a sequence of two observations and the entropy of a virtual distribution consistent with the market efficiency, according to which the conditional probabilities $\pi_{1,h}$ and $\pi_{0,h}$ are equal to $1/2$. This leads to the market information:
\begin{equation}\label{eq:MarketInfo}
I_h=-\sum_{i=1}^{2}{p_{i,h}\log_2\left(\frac{p_{i,h}}{2}\right)} + \sum_{i=1}^{2}{\left(p_{i,h}\pi_{i,h}\log_2\left(p_{i,h}\pi_{i,h}\right)+p_{i,h}(1-\pi_{i,h})\log_2\left(p_{i,h}(1-\pi_{i,h})\right)\right)}.
\end{equation}
Figure~\ref{fig:MarketInfoBTC} displays this quantity for each year for the Bitcoin, along with confidence intervals of the zero market information. These confidence intervals are those determined for an empirical distribution, that is for $h\rightarrow 0$~\cite{BG}. For small values of $h$, the Bitcoin seems to infringe the EMH. It is particularly the case in 2017, as well as, more softly, in 2016 and 2020. It confirms previous empirical results with the same method~\cite{BG} as well as with other approaches, with articles underlining the progressive decrease of Bitcoin inefficiency before 2018~\cite{DGM,KV19}.

We also observe in Figure~\ref{fig:MarketInfoBTC} the evolution of the market information with respect to the bandwidth. Above a certain threshold for $h$, the market information decreases when $h$ increases, leading to the conclusion, sometimes spurious, that the market is efficient. Once again, using a too big bandwidth may be misleading, regarding the interpretation in terms of market efficiency.

Last but not least, a popular approach in econophysics to determine whether the market is efficient is based on the Hurst exponent. The starting point for this last approach is the fractional Brownian motion (fBm)~\cite{MvN}, in which the Hurst exponent is a parameter $\mathcal H$ related to the covariance of the increments and thus to forecast~\cite{NP,Garcin2017} and to statistical arbitrages~\cite{GNR,GMR,GarcinForecast}. The market is supposed to be efficient for $\mathcal H$ close to $0.5$. Table~\ref{tab:HurstBTC} shows the Hurst exponent for the Bitcoin, calculated each year. The highest values are reached in 2017 and 2018. This method however does not depend on any bandwidth since the fBm assumes a Gaussian distribution. Many extensions of the fBm make it also possible to take into account other stylized facts of financial time series, such as non-Gaussian distributions~\cite{ST,WBMW,GarcinMPRE,AG}, time-varying parameters in a deterministic~\cite{BJR,PLV,ALV,Garcin2017,BP} or a stochastic fashion~\cite{AT,BPP,GarcinMPRE}, or even stationarity if one considers rates~\cite{FBA,GarcinLamperti,GarcinEstimLamp}. These numerous extensions of the fBm lead to other fruitful interpretations regarding the Hurst exponent and its link with market information~\cite{BG_fBm}. As a consequence, methodological choices, far beyond the sole question of the bandwidth selection, have a real impact on the analysis of a time series of prices and on the conclusion that the market is efficient or not.

\begin{table}[htbp]
\centering
\begin{tabular}{|l|c|}
\hline
Year & $\mathcal H$ \\
\hline
2015 & 0.470 \\
2016 & 0.498 \\
2017 & 0.528 \\
2018 & 0.537 \\
2019 & 0.506 \\
2020 & 0.509 \\
2021 & 0.494 \\
2022 & 0.503 \\
\hline
\end{tabular}
\begin{minipage}{0.9\textwidth}\caption{Hurst exponent $\mathcal H$ of the log-price of the BTC-USD, by year.}
\label{tab:HurstBTC}
\end{minipage}
\end{table}

\section{Conclusion}

The complexity of financial markets is at the heart of econophysical investigations. Since the distribution of price returns is a key element in quantitative finance, we have proposed a new method for defining and measuring the complexity of a probability distribution, in a financial context. Focusing on kernel densities, we have proposed a new method to select the key free parameter of kernel density estimation, namely the bandwidth. This method, which maximizes the complexity of the estimated density, is explicitly intended to avoid the two statistical pitfalls of overfitting and underfitting. We have compared our bandwidth selection methods with other classical approaches with the help of simulations. Finally, we have applied this method to real financial data. The conclusion of this empirical study is that, depending on the bandwidth or on other methodological choices, a same dataset may either be considered as consistent with the EMH or not. We can therefore never be too careful with the measurement tools of market efficiency and we recommend combining various approaches and not placing one's trust in a single statistic: analysing the complexity of a financial market requires a complex method.

\bibliographystyle{plain}

\bibliography{biblioKernelComplexity}



\end{document}